\documentclass[a4paper,11pt,oneside]{article}

\usepackage[english]{babel}
\usepackage[T1]{fontenc}
\usepackage[utf8]{inputenc}

\usepackage[pdftex,unicode]{hyperref}
\hypersetup{pdftitle=Streaming Approach to Quadratic Covariation Estimation Using Financial Ultra-High-Frequency Data}
\hypersetup{pdfauthor=Vladimír Holý and Petra Tomanová}

\usepackage{color}
\definecolor{myred}{rgb}{0.63,0.28,0.34}
\hypersetup{colorlinks=true, linkcolor=myred, anchorcolor=myred, citecolor=myred, filecolor=myred, urlcolor=myred}

\usepackage[margin=60pt]{geometry}
\usepackage{amsmath}
\usepackage{amssymb}
\usepackage{bm}

\usepackage[group-minimum-digits=3]{siunitx}

\usepackage{enumitem}

\usepackage{graphicx}
\usepackage{float}
\usepackage{hhline}
\usepackage{multirow}
\usepackage{rotating}
\usepackage{booktabs}

\usepackage[authoryear]{natbib}

\begin{document}

\begin{center}
{\Large \bfseries Streaming Approach to Quadratic Covariation Estimation \\ Using Financial Ultra-High-Frequency Data}
\end{center}

\begin{center}
{\bfseries Vladimír Holý} \\
University of Economics, Prague \\
Winston Churchill Square 4, 130 67 Prague 3, Czechia \\
\href{mailto:vladimir.holy@vse.cz}{vladimir.holy@vse.cz}
\end{center}

\begin{center}
{\bfseries Petra Tomanová} \\
University of Economics, Prague \\
Winston Churchill Square 4, 130 67 Prague 3, Czechia \\
\href{mailto:petra.tomanova@vse.cz}{petra.tomanova@vse.cz}
\end{center}

\begin{center}
{\itshape \today}
\end{center}

\noindent
\textbf{Abstract:}
We investigate the computational issues related to the memory size in the estimation of quadratic covariation, taking into account the specifics of financial ultra-high-frequency data. In multivariate price processes, we consider both contamination by the market microstructure noise and the non-synchronicity of the observations. We formulate a multi-scale, flat-top realized kernel, non-flat-top realized kernel, pre-averaging and modulated realized covariance estimators in quadratic form and fix their bandwidth parameter at a constant value. This allows us to operate with limited memory and formulate this  estimation as a streaming algorithm. We compare the performance of the estimators with fixed bandwidth parameter in a simulation study. We find that the estimators ensuring positive semidefiniteness require much higher bandwidth than the estimators without this constraint.
\\

\noindent
\textbf{Keywords:} Ultra-High-Frequency Data, Market Microstructure Noise, Quadratic Covariation, Streaming Algorithm.
\\

\noindent
\textbf{JEL Codes:} C32, C58, C63.
\\

\section{Introduction}
\label{sec:intro}

% high-frequency data
In finance, \cite{Engle2000} coined the term \emph{ultra-high-frequency data}, meaning irregularly spaced time series recorded at the highest possible frequency corresponding to each transaction or change in bid/ask offer. Financial high-frequency time series include stock prices, foreign exchange rates, and commodity prices. The availability of this high-frequency data allows econometricians to construct more precise models, while facing some new challenges.

% motivation
A key object in financial econometrics is volatility of the price process. A vast number of papers have demonstrated the benefits of using high-frequency data for volatility estimations in various financial applications. In derivative pricing, incorporating high-frequency information into pricing models of options leads to increased profits from option trading (see e.g., \citealp{Bandi2008b, Corsi2013, Sanfelici2014}). In risk management, using high-frequency estimates of volatility as a basis for determining the value-at-risk and expected shortfall increases their precision (see e.g., \citealp{Brownlees2010, Zikes2015, Bee2016}). In portfolio optimization, daily portfolio rebalancing based on high-frequency estimates of volatility improves portfolio allocations (see e.g., \citealp{Liu2009, Hautsch2015, Lunde2016}). Especially in the last case, covariance matrices of large dimensions are often required. Furthermore, in algorithmic trading, analyzing high-frequency data is necessary as decisions are also made at high frequencies (see e.g., \citealp{Christensen2012, Loveless2013, Arce2019}). Computationally efficient algorithms for estimating the variance and covariance are therefore crucial for practical use.

% quadratic variation
For continuous processes, volatility over a given time interval (e.g., a day) is typically measured by the quadratic variation (see e.g., \citealp{Andersen2001, Barndorff-Nielsen2002a}). In the theoretically ideal setting for the price process, it can be straightforwardly estimated by the realized variance. In practice, however, the so-called efficient price is concealed, as the observed prices are contaminated by market microstructure noise, making the realized variance significantly biased at high frequencies. Market microstructure noise is caused by various frictions in the trading process, such as the discretness of the prices, the bid--ask spread, and information effects (see e.g., \citealp{Hansen2006}).

% robust estimators
One way to deal with market microstructure noise is to sample the price process at lower frequencies and find the optimal trade-off between precision  and the bias due to this noise (see \citealp{Ait-Sahalia2005, Zhang2005, Bandi2008a}). A better way is to retain the entire dataset but use estimators that are robust to the noise. There are three dominant approaches in the non-parametric estimation of the quadratic variation: subsampling (see \citealp{Zhang2005, Zhang2006, Nolte2012, Ait-Sahalia2011}), autocovariance combining (see \citealp{Barndorff-Nielsen2008, Barndorff-Nielsen2009}) and pre-averaging (see \citealp{Jacod2009, Hautsch2013, Jacod2015}). These estimators have different motivations but very similar structure in the end. \cite{Sun2006} and \cite{Andersen2011} show that the multi-scale estimator of \cite{Zhang2006} based on subsampling and the flat-top realized kernel estimator of \cite{Barndorff-Nielsen2008} based on autocovariance combining can be expressed as a quadratic form. Furthermore, it can be shown that the pre-averaging estimator of \cite{Jacod2009} also has the structure of a quadratic form. Finally, when assuming a specific model for the price process, a parametric approach to estimating the volatility can also be adopted (see \citealp{Ait-Sahalia2005, Ait-Sahalia2010, Xiu2010, Holy2018d}).

% covariation
The estimation of the quadratic covariation between two price processes is even more challenging, due to the non-synchronicity of the trading. When the observations are simply synchronized using the previous-tick interpolation scheme, the Epps effect occurs and the realized covariance is biased (see e.g., \citealp{Hayashi2005, Zhang2011}). \cite{Zhang2011}, however, shows that subsampling in the two-scales estimator cancels not only the market microstructure noise but the Epps effect as well. \cite{Hayashi2005} propose a consistent estimator for the quadratic covariation using the original unaltered data. This approach is also adopted by \cite{Palandri2006}, \cite{Nolte2008}, \cite{Christensen2010} and \cite{Bibinger2011}. \cite{Martens2004}, \cite{Christensen2010} and \cite{Barndorff-Nielsen2011a} synchronize the observations using the refresh times of \cite{Harris1995}.

% our goal
\cite{Bandi2008b} highlight that a large portion of the high-frequency literature is devoted to the asymptotic properties of the volatility estimators, while the finite-sample properties are given less attention. Nevertheless, there are several studies comparing the finite-sample performance of volatility estimators according to various criteria. \cite{Liu2015} focus on estimation and forecasting accuracy, \cite{Gatheral2010} on statistical efficiency, implementation, and robustness, \cite{Bandi2008b} with \cite{Sanfelici2014} on their impact on option pricing, and \cite{Brownlees2010} on their impact on risk management. The goal of our paper is to compare the volatility estimators from yet another criterion -- computational efficiency with a focus on memory constraints.

% data stream
From a computational point of view, it is natural to consider financial high-frequency data as a data stream. A streaming algorithm can examine a sequence of inputs in a single pass only. The available memory for the computation is limited, and cannot store all the data. Note that different papers have used different definitions of this concept (see e.g., \citealp{Kontorovich2012} vs. \citealp{Cerny2019}). Related concepts are that of an online algorithm or a recursive algorithm, which focus on the updating scheme rather than memory constraints. Examples of streaming, online, and recursive algorithms in the field of econometrics include the estimation and diagnostics of the linear regression model by \cite{Cerny2019}, estimation of the ARMA model by \cite{Ouakasse2014}, estimation of the GARCH model by \cite{Aknouche2006}, estimation of the EWMA model by \cite{Hendrych2019}, estimation of the spot volatility by \cite{Dahlhaus2014}, and detection of changepoints by \cite{Bodenham2017}.

% our approach
In our paper, we focus on the non-parametric estimation of the quadratic covariation by means of streaming. We consider both the market microstructure noise and the non-synchronicity of the trading. First, we introduce the commonly used estimators that are robust to the market microstructure noise: the multi-scale estimator of \cite{Zhang2006}, the flat-top realized kernel estimator of \cite{Barndorff-Nielsen2008}, the non-flat-top realized kernel estimator of \cite{Barndorff-Nielsen2011a}, the pre-averaging estimator of \cite{Jacod2009}, and the modulated realized covariance estimator of \cite{Christensen2010}. With the aim of providing a unified and simple computational framework, we express the estimators as quadratic forms. All five estimators depend on a bandwidth parameter. These papers show that the optimal value of this parameter depends on the number of observations. Nevertheless, we assume the bandwidth parameter to be constant. This of course leads to sub-optimal performance but allows us to adopt a streaming algorithm with fixed memory, leading to fast computation. In the case of non-synchronous trading, we synchronize the observations by the refresh times method of \cite{Harris1995} with no computational issues.

% results
The main contributions of our paper are the following.
First, we express the quadratic covariation estimators as quadratic forms and illustrate the differences in their structure. Second, we propose using a streaming algorithm with fixed bandwidth for estimating the quadratic covariation when the computational performance is an issue. Third, we compare the quadratic covariation estimators and show the impact of the constant bandwidth parameter in a simulation study. We find that the multi-scale estimator is the most precise estimator in the case of small bandwidth, followed by the flat-top realized kernel estimator and the pre-averaging estimator. On the other hand, the non-flat-top realized kernel estimator and the modulated realized covariance estimator require much higher bandwidth but ensure positive semidefiniteness.

% structure
The rest of this paper is organized as follows. In Section \ref{sec:frame}, we present the standard framework for the price process and quadratic variation. In Section \ref{sec:est}, we describe the class of quadratic estimators and our streaming approach. In Section \ref{sec:sim}, we evaluate the performance of the estimators with fixed bandwidth using simulations. We conclude the paper in Section \ref{sec:con}.

\section{Theoretical Framework}
\label{sec:frame}

\subsection{Efficient Price Process}
\label{sec:frameEff}

We use the standard framework for the price process. Denote the $m$-dimensional logarithmic efficient price of a security by $P_t$ in continuous time $t \geq 0$. In theory, the efficient price is the true price of a security. In practice, however, the efficient price is unobservable. We assume that the efficient price follows a multivariate continuous Itô semimartingale given by
\begin{equation}
\label{eq:priceEff}
P_t = P_0 + \int_0^t \mu_s \, \mathrm{d} s + \int_0^t \sigma_s \, \mathrm{d} W_{s},
\end{equation}
where $\mu_s$ is a multivariate finite variation càdlàg drift process, $\sigma_s$ is a multivariate adapted càdlàg volatility process, and $W_s$ is a vector of independent Wiener processes. Note that both $\mu_s$ and $\sigma_s$ can vary over time and this class is therefore quite general.
However, a limitation is that we do not consider jumps, as the process is defined to have continuous paths.

\subsection{Quadratic Covariation}
\label{sec:frameQuadratic}

Our aim is to measure the volatility of the efficient price process over a given time interval. Without loss of generality, we may restrict ourselves to the time interval $[0, 1]$. The \emph{quadratic covariation} of the process $P_t$ on $[0, 1]$ is then given by
\begin{equation}
\label{eq:varQuad}
QV = \underset{\Delta_n \to 0}{\mathrm{plim}} \ \sum_{i=1}^n \Big( P_{T_i} - P_{T_{i-1}} \Big) \Big( P_{T_i} - P_{T_{i-1}} \Big)',
\end{equation}
where $\mathrm{plim}$ denotes the limit in probability and $\Delta_n = \max \{ T_1-T_0, T_2-T_1, \ldots, T_n-T_{n-1} \}$ is the maximal lag between observations of synchronized partitions $0 = T_0 < T_1 < \cdots < T_n = 1$.
As $\Delta_n \to 0$, we have that $n \to \infty$.
In our case of a continuous Itô semimartingale, the quadratic variation is identical to the \emph{integrated covariance} given by
\begin{equation}
\label{eq:varInt}
IV = \int_0^1 \sigma_s \sigma_s' \, \mathrm{d} s.
\end{equation}
For general semimartingales, however, they differ, due to the jump component.

\subsection{Observed Price Process}
\label{sec:frameObs}

As noted above, the efficient price is latent. This is for two reasons. First, we cannot observe the process in continuous time, but only at discrete points. Second, the efficient price is concealed due to various microstructural mechanisms, including the discretness of the prices, the bid--ask spread, and information effects. We capture this deviation from the efficient price by an additive noise term, referred to in the literature as the \emph{market microstructure noise}.

Assume that we observe an $m$-dimensional price process contaminated by the market microstructure noise at non-synchronous discrete times.
Denote the $k$-th univariate observed price process by $X_i^k$ at discrete times $0 \leq T_0^k < T_1^k < \cdots < T_{n^k}^k \leq 1$.
Note that the observations can be irregularly spaced. The $k$-th component of the latent price process $P_t^k$ and the observed price process $X_i^k$ are then related by
\begin{equation}
\label{eq:priceObs}
X_i^k = P_{T_i^k}^k + E_i^k, \quad E_i^k \sim (0, \omega^k) \quad \text{for } i = 0, \ldots, n^k,
\end{equation}
where $E_i^k$ is the market microstructure noise; it has expected value zero and standard deviation $\omega^k$. At this point, we do not impose additional assumptions on the noise $E_i^k$, as different estimators require different assumptions. We refer to the  papers for the individual estimators for more details. In an empirical study of financial markets, \cite{Hansen2006} show that the market microstructure noise is dependent both on its past values and the efficient price process.

\subsection{Data Synchronization}
\label{sec:frameSync}

Recall that we allow different univariate price processes to be observed at different times.
In the multivariate setting, this naturally creates difficulties.
An elegant approach is to synchronize the observation times, as proposed by \cite{Harris1995}.
We define the \emph{refresh times} $0 \leq T_0 < T_1 < \cdots < T_{n} \leq 1$ in the following way.
Let the initial refresh time be
\begin{equation}
T_0 = \max \left\{ T_0^1, \ldots, T_0^m \right\}.
\end{equation}
Next, let the subsequent refresh times be
\begin{equation}
T_i = \min \left\{ t : t \geq T_{j^1}^1 > T_{i-1}, t \geq T_{j^2}^2 > T_{i-1}, \ldots, t \geq T_{j^m}^m > T_{i-1} \right\} \quad \text{for } i = 1, \ldots, n.
\end{equation}
\cite{Martens2004} uses this scheme for the realized covariance.
\cite{Barndorff-Nielsen2011a} shows that this synchronization leads to consistent estimates of the quadratic covariation by the realized kernel estimator. \cite{Christensen2010} present similar results for the modulated realized covariance estimator.

After the synchronization, we can write the $m$-dimensional observed price process as $X_i = (X_i^1, X_i^2, \ldots, X_i^m)'$ and the market microstructure noise as $E_i = (E_i^1, E_i^2, \ldots, E_i^m)'$ with synchronized times $T_i$, $i = 0,\ldots,n$.
We define the observed returns as $Y_i = X_i - X_{i-1}$, $i = 1,\ldots,n$.
Finally, let $Y = (Y_1, \ldots, Y_n) = (Y_i^k)_{k=1,i=1}^{m,n}$ denote the matrix of the observed returns with rows corresponding to the assets and columns corresponding to the time.
Of course, not all prices are observed exactly at a moment that corresponds to a refresh time.
Often, only a single asset  has a new price observed.
For the other assets, the last observed price is used.
This is similar to the previous-tick approach, but the price interpolation is performed only at refresh times instead of all observation times.

\section{Streaming Estimation of Quadratic Variation}
\label{sec:est}

\subsection{Class of Quadratic Estimators}
\label{sec:class}

We estimate the quadratic covariation $QV$ in the presence of market microstructure noise by various non-parametric methods within a unified framework based on a quadratic form. \cite{Sun2006} and \cite{Andersen2011} consider the class of quadratic estimators in the univariate case of quadratic variation. In the multivariate case, the class of \textit{quadratic estimators} of the quadratic covariation based on the returns $Y_i$ are those having the form
\begin{equation}
\label{eq:quadReturns}
QE = \sum_{i=1}^n \sum_{j=1}^n w_{i,j} Y_i Y_j' = Y W Y',
\end{equation}
where $W = (w_{i,j})_{i=1,j=1}^{n,n}$ are weights associated with a given estimator. The formula can also be rewritten in terms of the actual prices $X_i$ as 
\begin{equation}
\label{eq:quadPrices}
QE = \sum_{i=0}^n \sum_{j=0}^n v_{i,j} X_i X_j' = X V X',
\end{equation}
where the weights $V = (v_{i,j})_{i=0,j=0}^{n,n}$ are given by $V = F'WF$. The elements of $F = (f_{i,j})_{i=1,j=0}^{n,n}$ are given by
\begin{equation}
\label{eq:quadTrans}
f_{i,j} = \left\{
\begin{array}{ll}
 1 & \text{for } j=i+1, \\
-1 & \text{for } j=i, \\
 0 & \text{otherwise.}
\end{array}
\right.
\end{equation}

\subsection{Streaming Approach}
\label{sec:stream}

In general, the computation of a quadratic form cannot be formulated as a streaming algorithm, as each observation is required to be stored in the memory. However, we can impose restrictions on the weight matrix $W$ in order to make the computation streaming. Assume the elements of the weight matrix $W$ satisfy
\begin{equation}
\label{eq:quadDiag}
w_{i,j} = \left\{
\begin{array}{ll}
u_{|i - j|} & \text{for } |i-j| < h, \\
0 & \text{otherwise},
\end{array}
\right.
\end{equation}
where $u = (u_0, \ldots, u_{h-1})'$ is the \emph{updating vector} determining a given streaming estimator. The matrix $W$ is therefore a symmetric $(2h - 1)$-diagonal matrix in which the elements in the main diagonal and each lower and upper diagonal are the same and determined by the updating vector $u$. At time $T_i$, the quadratic estimator can then be recursively computed by
\begin{equation}
\label{eq:quadStream}
QE_i = QE_{i-1} + u_0 Y_i Y_i' + \sum_{j=1}^{h-1} u_j Y_i Y_{i - j}' + \sum_{j=1}^{h-1} u_j Y_{i - j} Y_i'.
\end{equation}
Besides the previous matrix $QE_{i-1}$ and the vector of the current observations $Y_i$, we need to store the vectors of the previous observations $Y_{i-1}, \ldots, Y_{i-h+1}$ in memory. In total, that makes $m(m+h)$ real numbers.

In the following sections, we show that many estimators of the quadratic covariation can be expressed in a quadratic form \eqref{eq:quadReturns} satisfying the restriction \eqref{eq:quadDiag} if the bandwidth of the estimators is fixed at $h$ and possible edge effects are neglected.
On the one hand, a fixed bandwidth leads to sub-optimal performance of the estimators. On the other hand, it allows estimating the quadratic variation in the presence of market microstructure noise by means of a streaming algorithm. We define the \emph{edge effects} as the deviations of the weights $w_{ij}$ in the left upper corner $i,j = 1, \ldots, h-1$ and the right lower corner $i,j= n-h+2, \ldots,n$ from the values suggested by the updating vector $u$. The multi-scale, pre-averaging and modulated realized covariation estimators possess edge effects (see Figures \ref{fig:quadraticScale}, \ref{fig:quadraticPre} and \ref{fig:quadraticModul} respectively) while both the flat-top and non-flat-top realized kernel estimators do not (see Figures \ref{fig:quadraticFlat} and \ref{fig:quadraticKernel} respectively). \cite{Jacod2009} argue that the the edge effects are asymptotically unbiased in the case of the pre-averaging estimator and allow a simpler central limit theorem than in the case of the flat-top realized kernel estimator. In our streaming application, however, it is more suitable to omit the edge effects and modify all considered estimators in the fashion of \cite{Barndorff-Nielsen2008}.

\subsection{Realized Covariation}
\label{sec:rv}

\begin{figure}
\begin{center}
\includegraphics[width=0.9\textwidth]{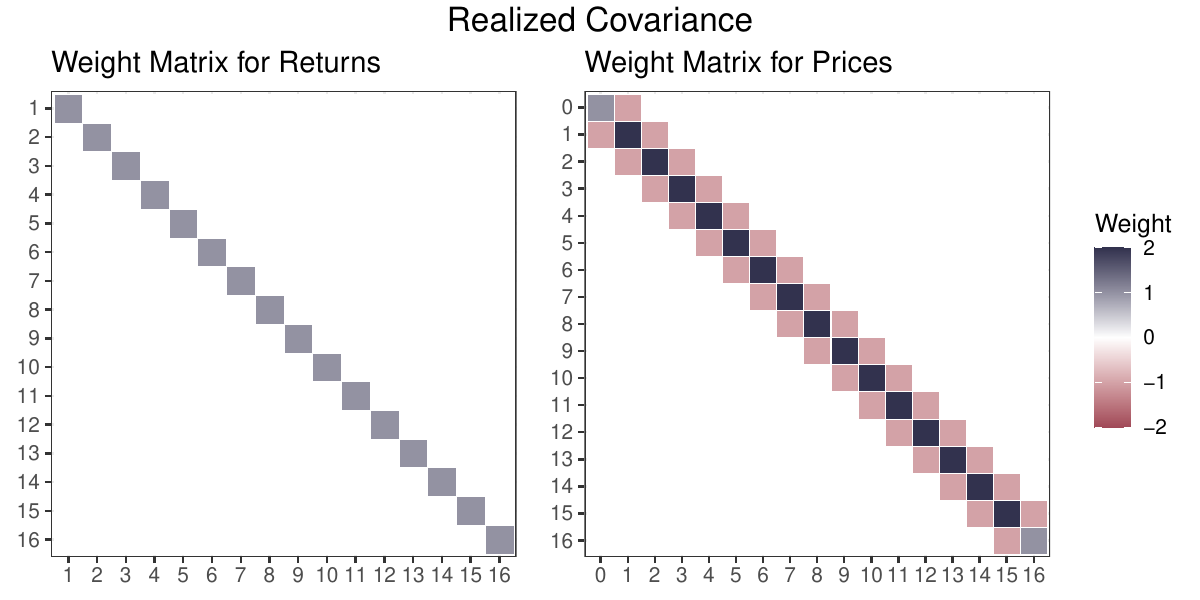}
\caption{Quadratic form of the realized variance with $n=16$.}
\label{fig:quadraticRealized}
\end{center}
\end{figure}

A natural estimator of the quadratic covariation is the \emph{realized covariance}, defined by
\begin{equation}
\label{eq:estRealized}
RV_n = \sum_{i=1}^n Y_i Y_i'.
\end{equation}
In the absence of noise, it is a consistent estimator of the quadratic covariation.
In the presence of noise, however, it is biased and inconsistent.
Note that in the asymptotics as $n \to \infty$, the time interval remains $[0, 1]$ but the frequency of observations increases.
The realized covariance can be easily expressed as a quadratic estimator using the weight matrix $W^{RV,n} = I$.
An example of this weight matrix is shown in Figure \ref{fig:quadraticRealized}.
The updating vector is simply $u^{RV,n} = (1,0, \ldots, 0)'$.

\subsection{Multi-Scale Estimator}
\label{sec:mse}

The first unbiased and consistent non-parametric estimator of the quadratic variation proposed in the literature is the two-scale estimator of \cite{Zhang2005}. It combines the average of realized variances computed at a lower frequency with realized variance at the highest possible frequency. The first term serves as a biased estimate of quadratic variation while the second term estimates the variance of the noise and therefore functions as the bias correction. \cite{Zhang2006} further generalizes the two-scale estimator to a multi-scale estimator based on the average realized variances computed at multiple frequencies. \cite{Ait-Sahalia2011} shows that the multi-scale estimator is robust to serial dependency in the market microstructure noise. In the subsampling spirit, \cite{Nolte2012} propose combining average realized variances using least squares. Extensions to the estimation of quadratic covariation have been proposed by \cite{Palandri2006}, \cite{Nolte2008}, \cite{Zhang2011} and \cite{Bibinger2011}.

\begin{figure}
\begin{center}
\includegraphics[width=0.9\textwidth]{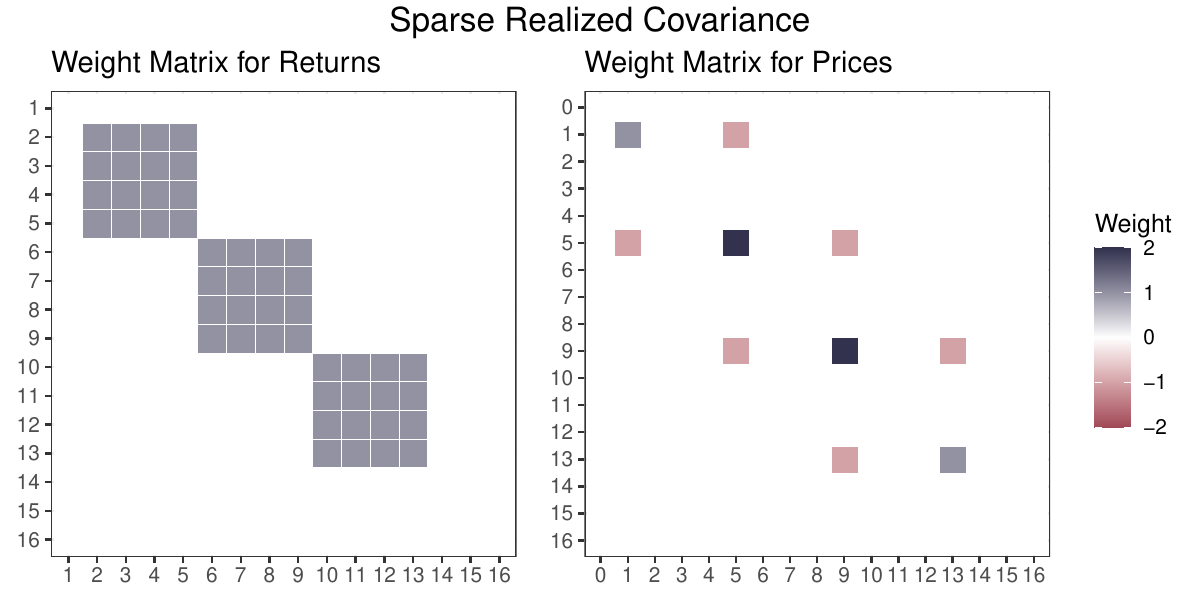}
\caption{Quadratic form of the sparse realized variance with $n=16$, $l=1$ and $s=4$.}
\label{fig:quadraticSparse}
\end{center}
\end{figure}

Before presenting the multi-scale estimator, we need to define some preliminary quantities. First, we introduce the sparse realized covariance, which is simply the realized covariance at a lower frequency. Let $l$ denote the index of the initial observation and $s$ denote the sampling interval. For example $l=1$ and $s=4$ would correspond to observations at times $\{T_1, T_5, T_9, T_{13}, \ldots \}$. The number of observed prices used in the estimation is then $\lfloor (n-l)/s \rfloor + 1$, where $\lfloor \cdot \rfloor$ denotes rounding down. The \textit{sparse realized covariance} is defined by
\begin{equation}
\label{eq:estSparse}
SRV_{n,l,s} = \sum_{k = 1}^{\lfloor (n-l)/s \rfloor} \left( X_{ks + l} - X_{(k - 1)s + l} \right)^2 = \sum_{k = 1}^{\lfloor (n-l)/s \rfloor} \left(\sum_{j = 1}^s Y_{(k-1)s+l+j} \right)^2.
\end{equation}
It can be expressed as a quadratic estimator using the weight matrix $W^{SRV,n,l,s}$ given by
\begin{equation}
\label{eq:weightSparse}
w_{i,j}^{SRV, n, l, s} = \left\{
\begin{array}{ll}
1 & \text{for } i,j \in [(k-1)s + l + 1 , ks + l], \  k=1,\ldots,\lfloor (n-l)/s \rfloor, \\
0 & \text{otherwise.}
\end{array}
\right.
\end{equation}
Figure \ref{fig:quadraticSparse} illustrates the quadratic form.
However, the sparse realized covariance cannot be expressed using an updating vector as it does not satisfy the requirements given by \eqref{eq:quadDiag}.

\begin{figure}
\begin{center}
\includegraphics[width=0.9\textwidth]{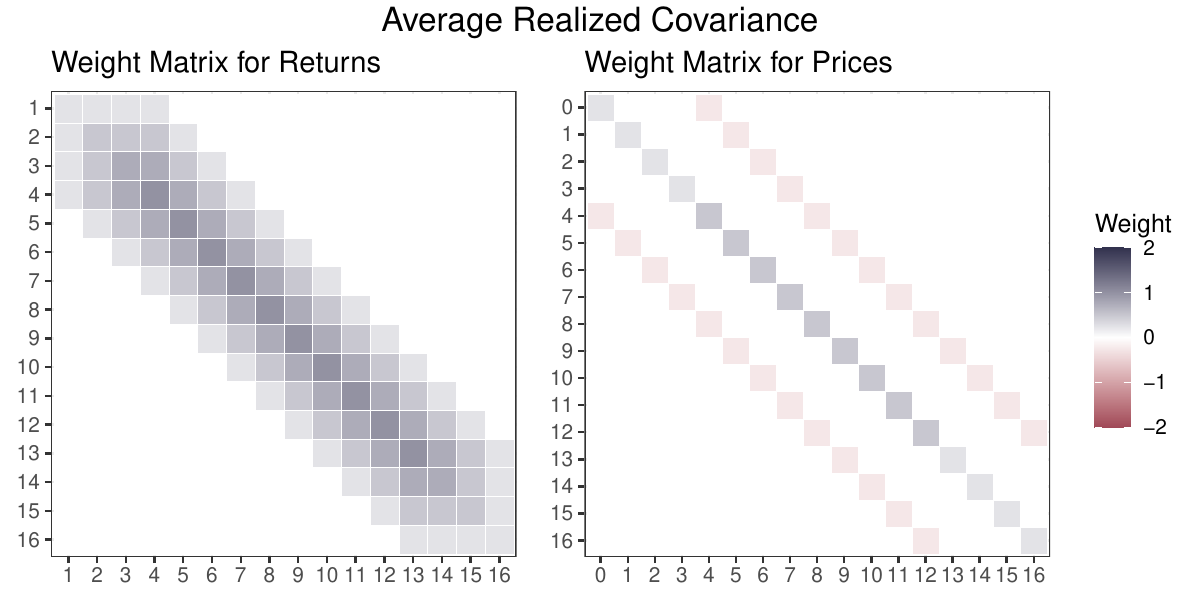}
\caption{Quadratic form of the average realized variance with $n=16$ and $s=4$.}
\label{fig:quadraticAverage}
\end{center}
\end{figure}

Next, we introduce the average realized covariance.
As the sparse realized covariance uses only a fraction of the available observations, it is natural to use all observations by averaging the sparse realized covariances over subgrids given by different $l$.
For a given sampling interval $s$, the \textit{average realized covariance} is defined by
\begin{equation}
\label{eq:estAverage}
ARV_{n,s} = \frac{1}{s} \sum_{l=0}^{s-1} SRV_{n,l,s}.
\end{equation}
Although this approach reduces the impact of the market microstructure noise, the average realized variance is still a biased estimator of the quadratic variation.
It can be expressed as a quadratic estimator using the weight matrix
\begin{equation}
\label{eq:weightAverage}
W^{ARV,n,s} = \frac{1}{s} \sum_{l=0}^{s-1} W^{SRV,n,l,s}.
\end{equation}
This weight matrix is shown in Figure \ref{fig:quadraticAverage}.
The updating vector $u^{ARV,n,s}$, omitting edge effects, is given by
\begin{equation}
\label{eq:updateAverage}
u_i^{ARV,n,s} = \frac{s - i}{s} \quad \text{for } i = 0, \ldots, s-1.
\end{equation}

\begin{figure}
\begin{center}
\includegraphics[width=0.9\textwidth]{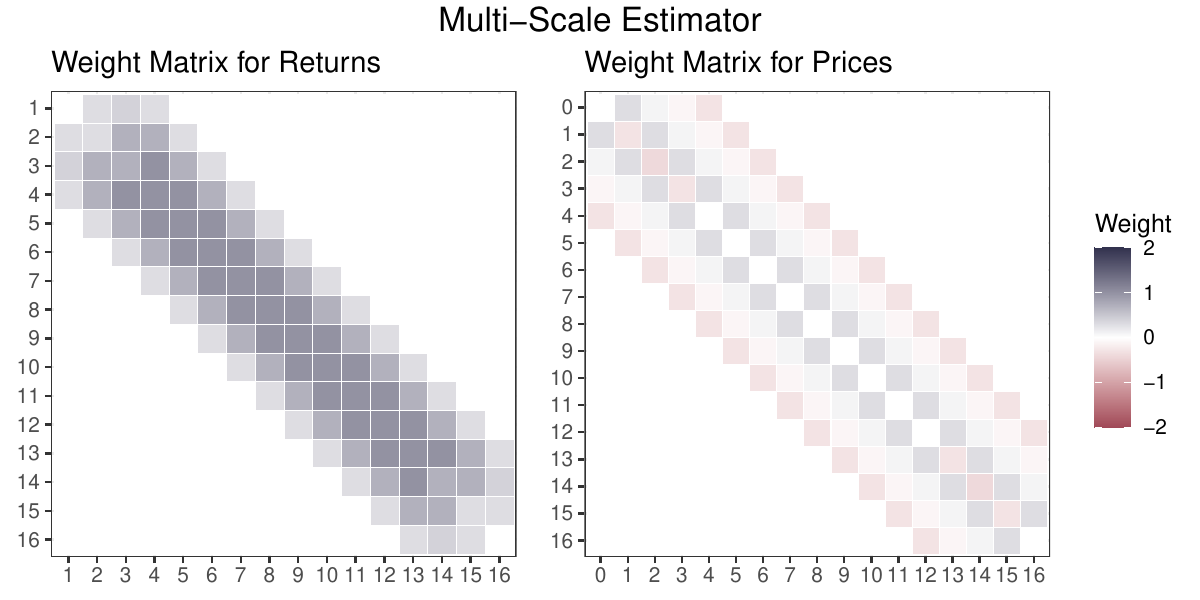}
\caption{Quadratic form of the multi-scale estimator with $n=16$ and $h=4$.}
\label{fig:quadraticScale}
\end{center}
\end{figure}

Finally, we present the \emph{multi-scale estimator} of \cite{Zhang2006}. It is a weighted average of the average realized variances based on the sampling intervals ranging from 1 up to the bandwidth $h$. It is given by
\begin{equation}
\label{eq:estScale}
MSE_{n,h} = \sum_{s=1}^h A(s,h) ARV_{n,s},
\end{equation}
where $A(s,h)$ is a weight function.
\cite{Zhang2006} suggests using
\begin{equation}
\label{eq:funScale}
A(s,h) = \frac{12s^2}{h^3 - h} - \frac{6s}{h^2-h}.
\end{equation}
The weight matrix is given by
\begin{equation}
\label{eq:weightScale}
W^{MSE,n,h} = \sum_{s=1}^h A(s,h) W^{ARV,n,s}.
\end{equation}
It is illustrated in Figure \ref{fig:quadraticScale}.
The updating vector $u^{MSE,n,h}$ omitting edge effects is given by
\begin{equation}
\label{eq:updateScale}
u_i^{MSE,n,h} = \sum_{s = i + 1}^h \frac{s - i}{s} A(s,h) \quad \text{for } i = 0, \ldots, h-1.
\end{equation}

\subsection{Realized Kernel Estimator}
\label{sec:rke}

Another approach for the robust estimation of the quadratic variation is combining realized autocovariances using kernel functions.
\cite{Barndorff-Nielsen2008} propose using flat-top kernels with unit weights at lags 0 and 1. \cite{Barndorff-Nielsen2009} use non-flat-top kernels with unit weight only at lag 0. Flat-top realized kernels provide a faster convergence rate but do not guarantee the non-negativity of the estimates. In contrast, non-flat-top realized kernels have sub-optimal convergence rates but ensure non-negativity. \cite{Barndorff-Nielsen2011a} propose non-flat-top kernels for the estimation of the quadratic covariation.

\begin{figure}
\begin{center}
\includegraphics[width=0.9\textwidth]{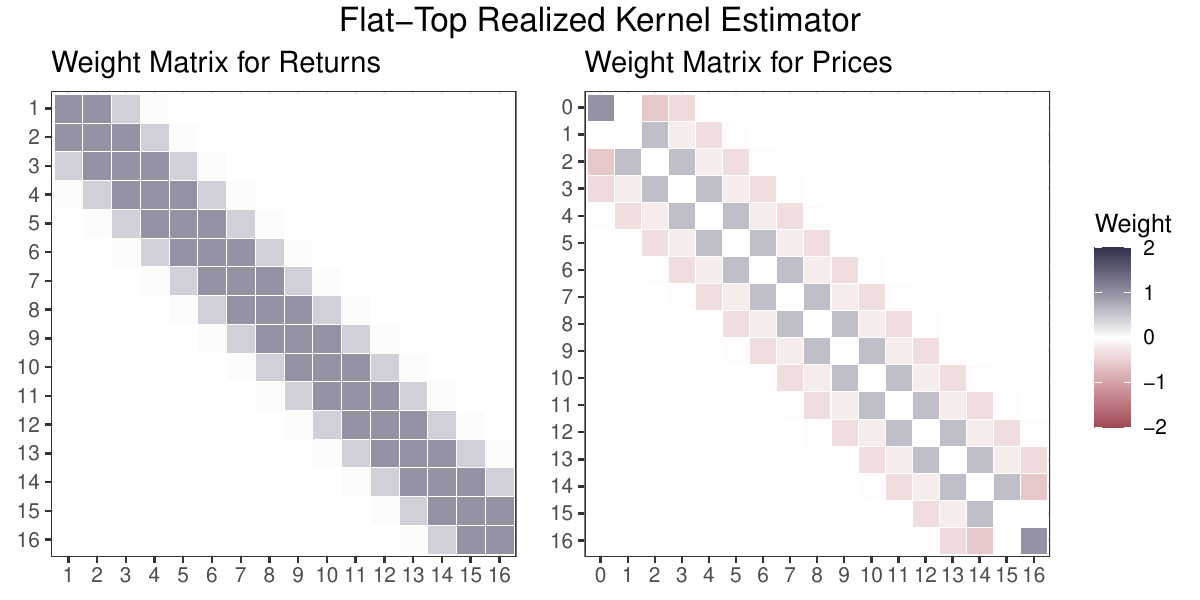}
\caption{Quadratic form of the flat-top realized kernel estimator with $n=16$ and $h=4$.}
\label{fig:quadraticFlat}
\end{center}
\end{figure}

The \emph{flat-top realized kernel estimator} of \cite{Barndorff-Nielsen2008} is defined by
\begin{equation}
\label{eq:estFlat}
RKE_{n,h} = RV_{n} + \sum_{j=1}^{h-1} K \left( \frac{j - 1}{h - 1} \right) \left( RA_{n,j} + RA_{n,-j} \right),
\end{equation}
where $K(\cdot)$ is a kernel function and $RA_{n,l}$ is the \emph{realized autocovariance} defined by
\begin{equation}
\label{eq:estAuto}
RA_{n,l} = \sum_{i=l+1}^{n} Y_{i} Y_{i-l}' \quad \text{for } l \geq 0,
\end{equation}
and $RA_{n,-l}'$ for $l < 0$.
\cite{Barndorff-Nielsen2008} consider many kernel functions and find that the \emph{modified Tukey--Hanning kernel of order 2} is near efficient.
It is defined by
\begin{equation}
\label{eq:funTukey}
K(x) =  \sin^2 \left( \frac{\pi}{2} (1 - x)^2 \right).
\end{equation}
The flat-top realized kernel estimator can be expressed as a quadratic form with weight matrix $W^{RKE,n,h}$ given by
\begin{equation}
\label{eq:weightFlat}
w_{i,j}^{RKE,n,h} = \left\{
\begin{array}{ll}
1 & \text{for } i=j, \\
K \left( \frac{|i-j| - 1}{h - 1} \right) & \text{for } 1 \leq |i-j| < h, \\
0 & \text{otherwise.}
\end{array}
\right.
\end{equation}
It is illustrated in Figure \ref{fig:quadraticFlat}.
The updating vector $u^{RKE,n,h}$ is given by
\begin{equation}
\label{eq:updateFlat}
u_{i}^{RKE,n,h} = \left\{
\begin{array}{ll}
1 & \text{for } i=0, \\
K \left( \frac{i - 1}{h - 1} \right) & \text{for } i = 1, \ldots, h-1. \\
\end{array}
\right.
\end{equation}

\begin{figure}
\begin{center}
\includegraphics[width=0.9\textwidth]{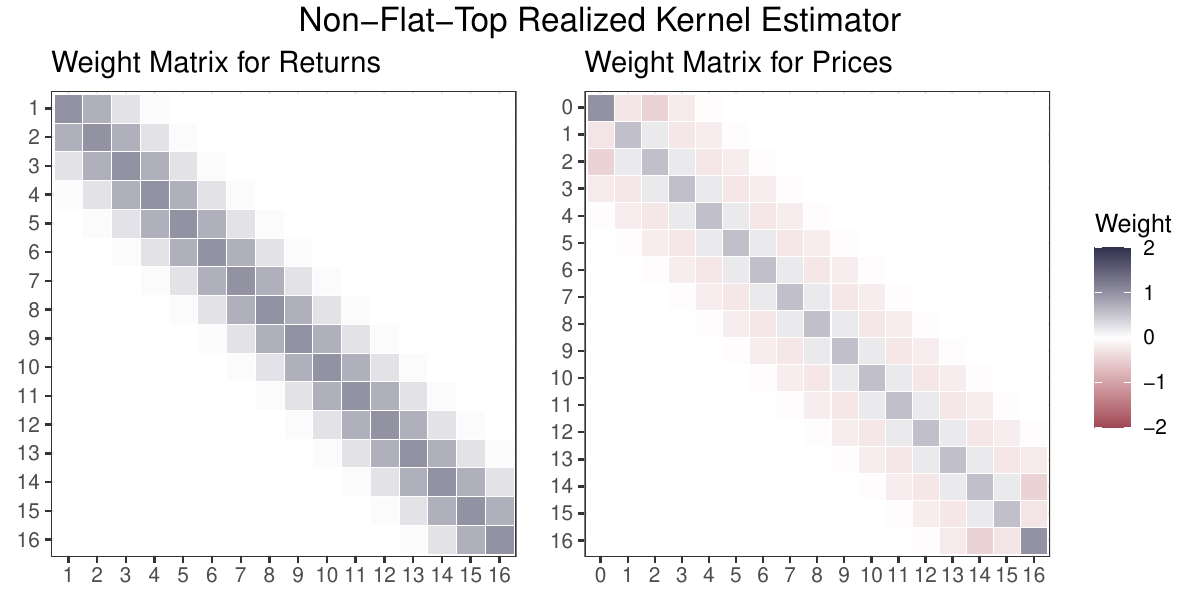}
\caption{Quadratic form of the non-flat-top realized kernel estimator with $n=16$ and $h=4$.}
\label{fig:quadraticKernel}
\end{center}
\end{figure}

To ensure a positive semidefinite covariance matrix, the \emph{non-flat-top realized kernel} of \cite{Barndorff-Nielsen2011a} can be used.
It is defined by
\begin{equation}
\label{eq:estKernel}
PD\text{-}RKE_{n,h} = \sum_{j=-h+1}^{h-1} K \left( \frac{|j|}{h} \right) RA_{n,j}.
\end{equation}
Both \cite{Barndorff-Nielsen2009} in the univariate case and \cite{Barndorff-Nielsen2011a} in the multivariate case suggest using the \emph{Parzen kernel} given by
\begin{equation}
\label{eq:funParzen}
K(x) =  \left\{
\begin{array}{ll}
1 - 6x^2 + 6x^3 & \text{for } 0 \leq x < \frac{1}{2}, \\
2 (1 - x)^3 & \text{for } \frac{1}{2} \leq x \leq 1. \\
\end{array}
\right.
\end{equation}
The non-flat-top realized kernel estimator can be expressed as a quadratic form with weight matrix $W^{PD\text{-}RKE,n,h}$ given by
\begin{equation}
\label{eq:weightKernel}
w_{i,j}^{PD\text{-}RKE, n, h} = \left\{
\begin{array}{ll}
K \left( \frac{|i-j|}{h} \right) & \text{for } |i-j| < h, \\
0 & \text{otherwise.}
\end{array}
\right.
\end{equation}
It is illustrated in Figure \ref{fig:quadraticKernel}.
The updating vector $u^{PD\text{-}RKE,n,h}$ is simply given by
\begin{equation}
\label{eq:updateKernel}
u_{i}^{PD\text{-}RKE,n,h} = K \left( \frac{i}{h} \right) \quad \text{for } i = 0, \ldots, h-1. \\
\end{equation}

\subsection{Pre-Averaging Estimator}
\label{sec:pae}

The third class of estimators we present are the pre-averaging estimators pioneered by \cite{Jacod2009}.
The idea is to locally average the returns and then sum their squares.
\cite{Hautsch2013} extend the theory of pre-averaging estimators to accommodate a serial dependence in the market microstructure noise and jumps in the price process.
\cite{Jacod2015} propose an adaptive method for the choice of the bandwidth parameter.
\cite{Christensen2010} extend the pre-averaging estimator to the multivariate setting and use the name modulated realized covariance instead.

\begin{figure}
\begin{center}
\includegraphics[width=0.9\textwidth]{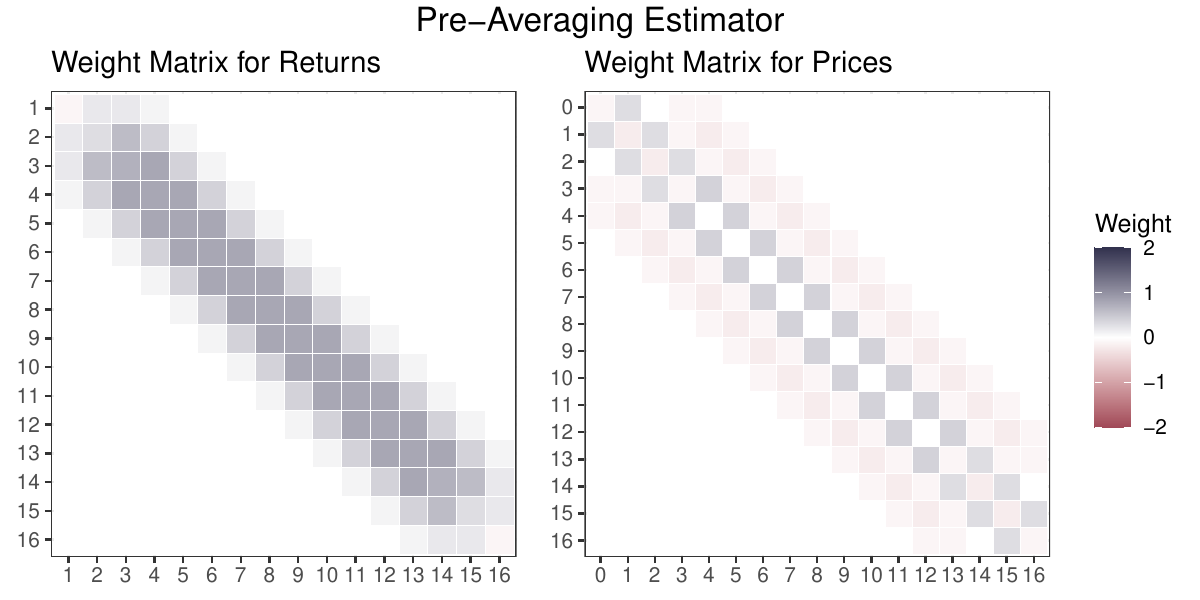}
\caption{Quadratic form of the pre-averaging estimator with $n=16$ and $h=4$.}
\label{fig:quadraticPre}
\end{center}
\end{figure}

The \emph{pre-averaging estimator} of \cite{Jacod2009} is based on the averaged returns, and is given by
\begin{equation}
\begin{aligned}
\bar{Y}_{i} &= \sum_{j=0}^{h-1} G \left( \frac{j+1}{h+1} \right) Y_{i+j}, \\
\end{aligned}
\end{equation}
where $G(\cdot)$ is a suitable weight function.
\cite{Jacod2009} suggest using
\begin{equation}
G(x) = \min \left\{ x, 1-x \right\}.
\end{equation}
The pre-averaging estimator is then defined by
\begin{equation}
\label{eq:estPre}
PAE_{n,h} = \frac{1}{\psi_2} \sum_{i = 1}^{n - h + 1} \bar{Y}_{i} \bar{Y}_{i}' - \frac{\psi_1}{2 \psi_2} RV_{n},
\end{equation}
where
\begin{equation}
\begin{aligned}
\psi_1 &= \sum_{j=0}^{h} \left( G \left( \frac{j+1}{h+1} \right) - G \left( \frac{j}{h+1} \right) \right)^2 , \\
\psi_2 &= \sum_{j=0}^{h-1} G \left( \frac{j + 1}{h + 1} \right)^2. \\
\end{aligned}
\end{equation}
The realized variance term serves as a correction for the bias.
Note that similarly to \cite{Jacod2015}, we use a simpler expression for the estimator than that introduced in \cite{Jacod2009}, and omit terms related to the asymptotics of the bandwidth parameter.
We also do not include the adjustment for the sample size, as the final number of observations is unknown in the typical streaming setting.
The pre-averaging estimator can be expressed as a quadratic estimator using the weight matrix $W^{PAE,n,h}$ given by
\begin{equation}
\label{eq:weightPre}
w^{PAE,n,h}_{i,j} = \left\{
\begin{array}{ll}
\frac{1}{\psi^2} \sum_{k=\max \{ 0, i+h-n-1 \} }^{\min \{ h-1, i-1 \} } G^2 \left( \frac{k+1}{h+1} \right) - \frac{\psi_1}{2 \psi_2} & \text{for } i=j, \\
\frac{1}{\psi^2} \sum_{k=\max \{ 0, i+h-n-1, j+h-n-1 \} }^{\min \{ h-1-|i-j|, i-1, j-1\} } G \left( \frac{k+1}{h+1} \right) G \left( \frac{k+1+|i-j|}{h+1} \right) & \text{for } 1 \leq |i-j| < h, \\
0 & \text{otherwise.}
\end{array}
\right.
\end{equation}
The weight matrix is illustrated in Figure \ref{fig:quadraticPre}.
The updating vector $u^{PAE,n,h}$, omitting the edge effects, is given by
\begin{equation}
\label{eq:updatePre}
u_i^{PAE,n,h} = \left\{
\begin{array}{ll}
\frac{1}{\psi^2} \sum_{j=0}^{h-1} G^2 \left( \frac{j+1}{h+1} \right) - \frac{\psi_1}{2 \psi_2} & \text{for } i=0, \\
\frac{1}{\psi^2} \sum_{j=0}^{h-1-i} G \left( \frac{j+1}{h+1} \right) G \left( \frac{j+1+i}{h+1} \right) & \text{for } i = 1, \ldots, h-1. \\
\end{array}
\right.
\end{equation}

\begin{figure}
\begin{center}
\includegraphics[width=0.9\textwidth]{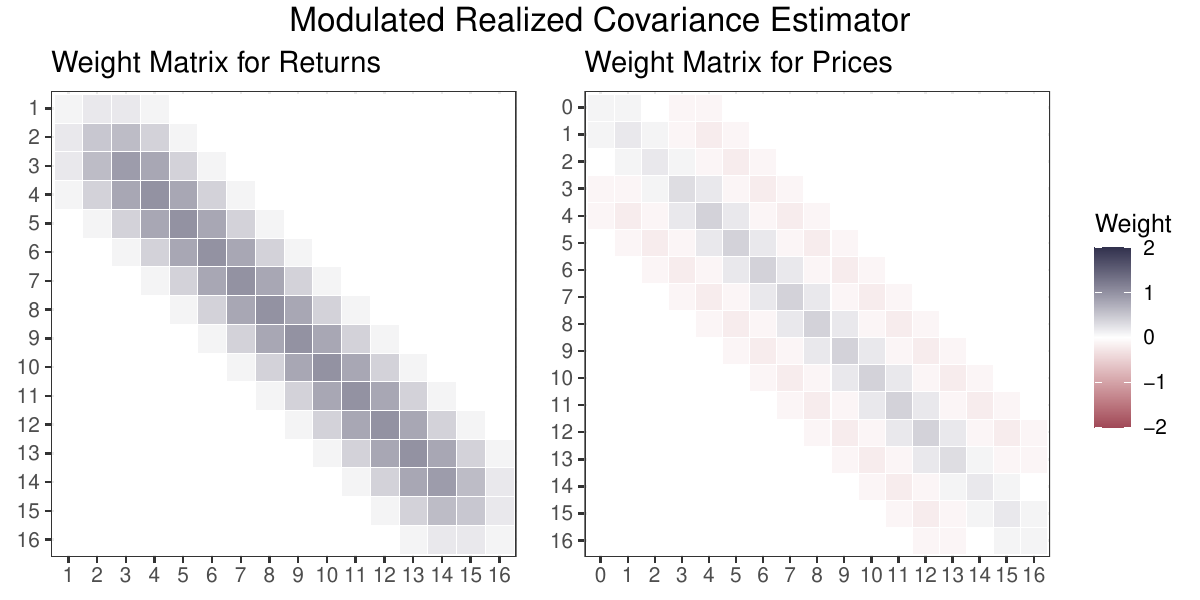}
\caption{Quadratic form of the modulated realized covariation estimator with $n=16$ and $h=4$.}
\label{fig:quadraticModul}
\end{center}
\end{figure}

When the bias correction term in \eqref{eq:estPre} is omitted, the resulting estimator is guaranteed to be positive semidefinite.
Similary to the non-flat-top realized kernel estimator, however, it has a sub-optimal convergence rate.
\cite{Christensen2010} define the \emph{modulated realized covariance estimator} by
\begin{equation}
\label{eq:estModul}
PD\text{-}PAE_{n,h} = \frac{1}{\psi_2} \sum_{i = 1}^{n - h + 1} \bar{Y}_{i} \bar{Y}_{i}'.
\end{equation}
The weight matrix $W^{PD\text{-}PAE,n,h}$ is given by
\begin{equation}
\label{eq:weightModul}
w^{PD\text{-}PAE,n,h}_{i,j} = \frac{1}{\psi^2} \sum_{k=\max \{ 0, i+h-n-1, j+h-n-1 \} }^{\min \{ h-1-|i-j|, i-1, j-1\} } G \left( \frac{k+1}{h+1} \right) G \left( \frac{k+1+|i-j|}{h+1} \right) \quad \text{for } 1 \leq |i-j| < h.
\end{equation}
This weight matrix is shown in Figure \ref{fig:quadraticModul}.
The updating vector $u^{PD\text{-}PAE,n,h}$, omitting the edge effects, is given by
\begin{equation}
\label{eq:updateModul}
u_i^{PD\text{-}PAE,n,h} = \frac{1}{\psi^2} \sum_{j=0}^{h-1-i} G \left( \frac{j+1}{h+1} \right) G \left( \frac{j+1+i}{h+1} \right) \quad \text{for } i = 0, \ldots, h-1. \\
\end{equation}

\section{Simulation Study}
\label{sec:sim}

\subsection{Setup for the Simulations}
\label{sec:simSetup}

To compare the finite-sample performance of the estimators, we conducted a simulation study. We consider the same model for the observed price process as \cite{Barndorff-Nielsen2011a}. The individual efficient price $P^k$, $k = 1,\ldots,m$, in continuous time follows
\begin{equation}
\label{eq:simEfficient}
\mathrm{d} P^k = \mu \mathrm{d}t + \mathrm{d} V^k + \mathrm{d} F^k,
\end{equation}
where $V^k$ and the $F^k$ are given by
\begin{equation}
\label{eq:simComponent}
\begin{aligned}
\mathrm{d} V^k &= \rho S^k \mathrm{d} B^k, \\
\mathrm{d} F^k &= \sqrt{1 - \rho^2} S^k \mathrm{d} B^0, \\
\end{aligned}
\end{equation}
and $B_0,B_1, \ldots, B_m$ are independent Wiener processes. The volatility process $S^k$ is given by
\begin{equation}
\label{eq:simVolatility}
\begin{aligned}
S^k &= \exp \left( \alpha + \beta U^k \right), \\
\mathrm{d} U^k &= \theta U^k \mathrm{d} t + \mathrm{d} B^k. \\
\end{aligned}
\end{equation}
We restrict ourselves to two-dimensional processes, i.e., $m=2$. We generate the observation times $T_i^k$ by two independent Poisson point processes with scale parameters $\lambda = (\lambda_1, \lambda_2)$. We standardize the time so that one unit corresponds to one second in a trading day that is 6.5 hours long. The time window for which we compute the quadratic covariation is therefore \num{23400} seconds long. In the case of two independent Poisson processes, the refresh time sampling results in an average number of observations of
\begin{equation}
n = \num{23400} \frac{\lambda_1 + \lambda_2}{\lambda_1^2 + \lambda_1 \lambda_2 + \lambda_2^2}.
\end{equation} 
We contaminate the efficient prices by market microstructure noise. The observed prices $X^k$ are then
\begin{equation}
\label{eq:simObserved}
X_i^k = P_{T_i^k}^k + E_i^k, \quad E_i^k \sim \mathrm{N} \left(0, \omega^k \right), \quad \omega^k = \xi \sqrt[4]{ \frac{1}{n+1} \sum_{i=0}^n \left( S^k_{T^k_i} \right)^4 }.
\end{equation}

The simulations were perfomed 1\,000\,000 times, i.e., we simulated 1\,000\,000 days. The volatility process \eqref{eq:simVolatility} was simulated using the exact simulation algorithm for the Ornstein--Uhlenbeck process. The first observation for each day was generated using the stationary distribution $Q_0^k \sim \mathrm{N} \left(0, (-2 \theta)^{-1} \right)$. The efficient price process \eqref{eq:simEfficient} with its components \eqref{eq:simComponent} was simulated using the Euler method. The initial observation was set to $P^k_0 = 0$.

As in \cite{Barndorff-Nielsen2011a}, we set the parameter values for the efficient prices to $\mu=0.03$, $\rho=-0.3$, $\alpha=-0.3125$, $\beta=0.125$, and $\theta=-0.025$. These values reflect the empirical properties of financial markets and result in the expected value of quadratic covariation given by
\begin{equation}
\mathrm{E} \left[ QV \right] = \begin{pmatrix}
1.00 & 0.67 \\
0.67 & 1.00
\end{pmatrix}.
\end{equation}
Furthermore, we consider three values for the noise: $\xi^2=0$ (referred to as None), $\xi^2=0.001$ (referred to as Small) and $\xi^2=0.01$ (referred to as Large). We consider two frequencies: $\delta = (1, 0.5)$ (referred to as Moderate) and $\delta = (0.1, 0.05)$ (referred to as High). We choose much higher frequencies than \cite{Barndorff-Nielsen2011a}, so as to reflect current trading behavior. The scenarios with moderate frequency have on average \num{23400} and \num{46800} observations, respectively, while the high frequency scenarios have \num{234000} and \num{468000} observations, respectively. After the refresh time sampling, we have on average \num{20057} observations when using the moderate frequency and \num{200571} observations when using the high frequency.

\subsection{Results}
\label{sec:simRes}

\begin{figure}
\begin{center}
\includegraphics[width=0.9\textwidth]{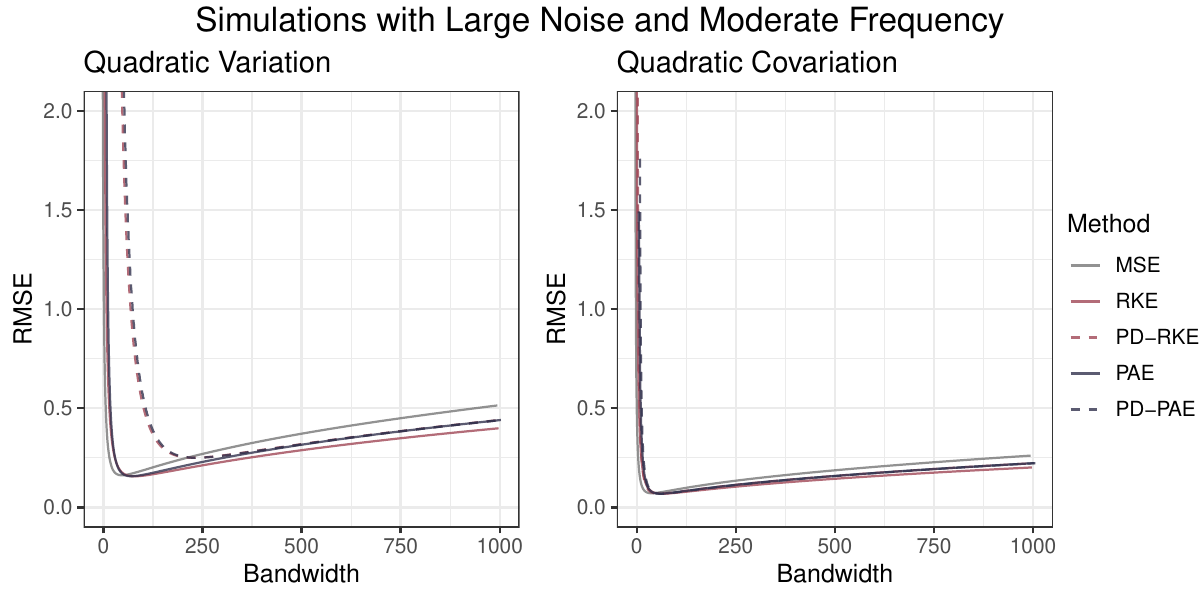}
\caption{Root-mean-square error of quadratic covariation estimates for various bandwidth parameters when $\xi^2=0.01$ and $\delta = (1, 0.5)$.}
\label{fig:simMod}
\end{center}
\end{figure}

\begin{figure}
\begin{center}
\includegraphics[width=0.9\textwidth]{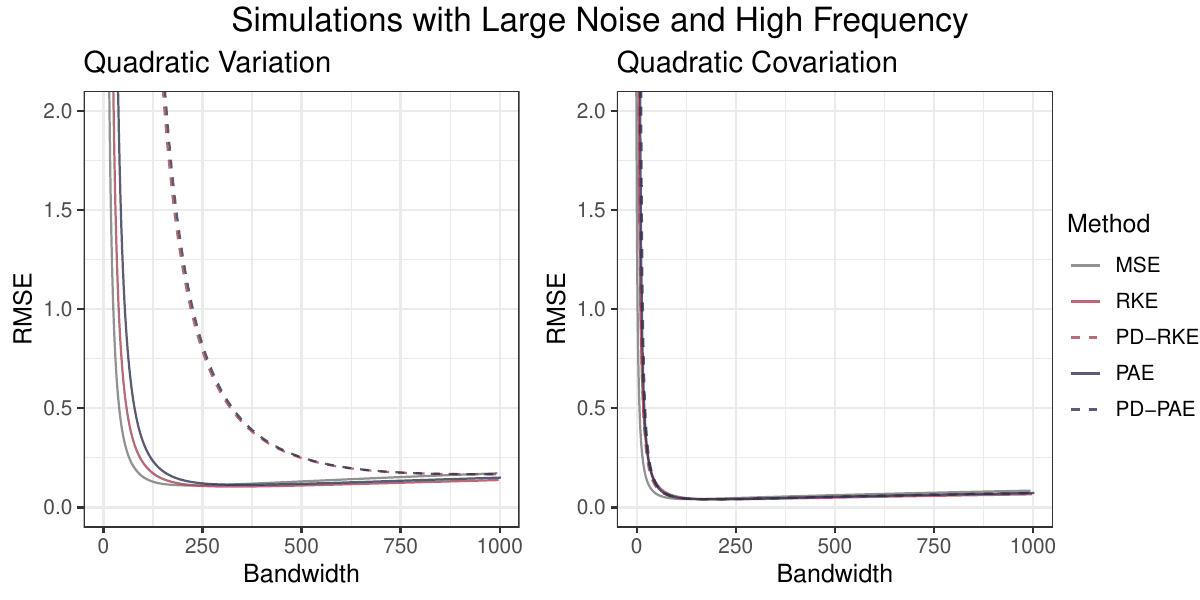}
\caption{Root-mean-square error of quadratic covariation estimates for various bandwidth parameters when $\xi^2=0.01$ and $\delta = (0.1, 0.05)$.}
\label{fig:simHigh}
\end{center}
\end{figure}

The results of the simulation study are presented in Table \ref{tab:sim}, Figure \ref{fig:simMod}, and Figure \ref{fig:simHigh}.
In Table \ref{tab:sim}, we choose the best performing bandwidth parameter according to the root-mean-square error.
In Figures \ref{fig:simMod} and \ref{fig:simHigh}, we investigate the behavior of the root-mean-square error for values of the bandwidth parameter ranging from 2 to \num{1000} (regardless of the number of observations).

The realized variance is the best estimator in the univariate case when there is no market microstructure noise.
When noise is present, however, the realized variance is significantly biased with increasing frequency.
In the multivariate case, the realized covariance is negatively biased under the refresh time sampling.
These are both standard results well covered in the literature.

The multi-scale, flat-top realized kernel and pre-averaging estimators perform comparably.
They have lower root-mean-square error of the univariate estimates than the non-flat-top realized kernel and modulated realized covariance estimators in the presence of market microstructure noise.
However, they do not ensure positive semidefiniteness, which is a major drawback.
Table \ref{tab:sim} with Figures \ref{fig:simMod} and \ref{fig:simHigh} show that the multi-scale estimator requires the lowest bandwidth for the optimal performance and is the best choice when the bandwidth is lower than optimal.
As this is the natural situation in the streaming setting, we recommend adopting the multi-scale estimator in practice when memory is limited.
When the bandwidth is higher than optimal, however, the flat-top realized kernel estimator is the most precise.
The pre-averaging estimator represents a middle way.

The non-flat-top realized kernel and modulated realized covariance estimators ensure positive semidefiniteness.
On the other hand, they are less precise and require higher bandwidth in the univariate case, as shown in Table \ref{tab:sim} and Figures \ref{fig:simMod} and \ref{fig:simHigh}.
In the case of the quadratic covariation between two series, however, they are quite comparable to the estimators not ensuring positive semidefiniteness.
Interestingly, the non-flat-top realized kernel estimator has almost the same performance as the  modulated realized covariance estimator, even though their structures differ, as illustrated in Figures \ref{fig:quadraticKernel} and \ref{fig:quadraticModul}.
As requiring a positive semidefinite covariance matrix is quite a reasonable requirement, either of these two methods should be used when the bandwidth parameter is not limited.

\begin{table}
\begin{center}
\begin{tabular}{lllrrrrrr}
\toprule
\multicolumn{3}{c}{Simulation Scenario} & \multicolumn{3}{c}{Quadratic Variation} & \multicolumn{3}{c}{Quadratic Covariation} \\
Noise & Freq. & Method & Band. & Bias & RMSE & Band. & Bias & RMSE \\
\midrule
  None  & Mod. & RV     &   1 &   -0.00 &    0.02 &   1 & -0.19 &  0.27 \\ 
  None  & Mod. & MSE    &   2 &   -0.00 &    0.04 &   2 & -0.00 &  0.02 \\ 
  None  & Mod. & RKE    &   2 &   -0.00 &    0.04 &   2 & -0.00 &  0.02 \\ 
  None  & Mod. & PD-RKE &   2 &   -0.00 &    0.03 &  11 & -0.01 &  0.03 \\ 
  None  & Mod. & PAE    &  18 &   -0.02 &    0.07 &  18 & -0.01 &  0.03 \\ 
  None  & Mod. & PD-PAE &   2 &   -0.00 &    0.03 &  10 & -0.01 &  0.03 \\ 
  None  & High & RV     &   1 &   -0.00 &    0.01 &   1 & -0.19 &  0.27 \\ 
  None  & High & MSE    &   2 &   -0.00 &    0.01 &   2 & -0.00 &  0.01 \\ 
  None  & High & RKE    &   2 &   -0.00 &    0.01 &   2 & -0.00 &  0.01 \\ 
  None  & High & PD-RKE &   2 &   -0.00 &    0.01 &  18 & -0.00 &  0.01 \\ 
  None  & High & PAE    &  28 &   -0.01 &    0.03 &  28 & -0.00 &  0.01 \\ 
  None  & High & PD-PAE &   2 &   -0.00 &    0.01 &  16 & -0.00 &  0.01 \\ 
  Small & Mod. & RV     &   1 &   40.25 &   75.13 &   1 & -0.19 &  0.44 \\ 
  Small & Mod. & MSE    &  17 &    0.00 &    0.09 &  14 & -0.00 &  0.04 \\ 
  Small & Mod. & RKE    &  26 &    0.00 &    0.09 &  22 & -0.00 &  0.04 \\ 
  Small & Mod. & PD-RKE &  85 &    0.04 &    0.14 &  18 & -0.00 &  0.04 \\ 
  Small & Mod. & PAE    &  22 &   -0.01 &    0.08 &  20 & -0.01 &  0.04 \\ 
  Small & Mod. & PD-PAE &  84 &    0.04 &    0.14 &  17 & -0.00 &  0.04 \\ 
  Small & High & RV     &   1 &  395.54 &  738.56 &   1 & -0.19 &  1.11 \\ 
  Small & High & MSE    &  75 &    0.01 &    0.05 &  45 & -0.00 &  0.02 \\ 
  Small & High & RKE    & 108 &    0.01 &    0.05 &  69 & -0.00 &  0.02 \\ 
  Small & High & PD-RKE & 348 &    0.02 &    0.09 &  57 & -0.00 &  0.02 \\ 
  Small & High & PAE    & 114 &    0.01 &    0.06 &  58 & -0.00 &  0.02 \\ 
  Small & High & PD-PAE & 346 &    0.02 &    0.09 &  57 & -0.00 &  0.02 \\ 
  Large & Mod. & RV     &   1 &  402.48 &  751.65 &   1 & -0.20 &  3.49 \\ 
  Large & Mod. & MSE    &  51 &    0.02 &    0.16 &  44 & -0.00 &  0.07 \\ 
  Large & Mod. & RKE    &  82 &    0.02 &    0.16 &  68 & -0.00 &  0.07 \\ 
  Large & Mod. & PD-RKE & 228 &    0.07 &    0.25 &  56 & -0.00 &  0.07 \\ 
  Large & Mod. & PAE    &  67 &    0.02 &    0.16 &  56 & -0.00 &  0.07 \\ 
  Large & Mod. & PD-PAE & 228 &    0.07 &    0.25 &  55 & -0.00 &  0.07 \\ 
  Large & High & RV     &   1 & \num{3955.40} & \num{7385.83} &   1 & -0.20 & 10.75 \\ 
  Large & High & MSE    & 232 &    0.03 &    0.11 & 137 & -0.00 &  0.04 \\ 
  Large & High & RKE    & 331 &    0.03 &    0.10 & 218 & -0.00 &  0.04 \\ 
  Large & High & PD-RKE & 958 &    0.05 &    0.17 & 181 & -0.00 &  0.04 \\ 
  Large & High & PAE    & 348 &    0.03 &    0.11 & 181 & -0.00 &  0.04 \\ 
  Large & High & PD-PAE & 956 &    0.05 &    0.17 & 179 & -0.00 &  0.04 \\   
\bottomrule
\end{tabular}
\caption{Bias and root-mean-square error of quadratic covariation estimates for the best performing bandwidth parameter in various combinations of noise and frequency.}
\label{tab:sim}
\end{center}
\end{table}

\section{Conclusion}
\label{sec:con}

We have dealt with the estimation of the quadratic covariation, taking into account the specifics of financial ultra-high-frequency data. Namely, we have considered both market microstructure noise and non-synchronous observations. To our knowledge, the high-frequency literature lacks a comprehensive overview of quadratic covariation estimators in a unified framework. We have remedied this and presented the multi-scale, flat-top realized kernel, non-flat-top realized kernel, pre-averaging and modulated realized covariance estimators as quadratic forms. We have also illustrated the differences in their structures. This is the first contribution of the paper.

We have approached the problem of estimating quadratic covariation from a computational point of view, focusing on limited memory. We have used the convenient quadratic structure and shown that the estimates can be computed by a streaming algorithm when the bandwith is fixed and edge effects are neglected. The streaming representation of the estimators is crucial, especially when the covariance matrix is large. This is the second contribution of the paper.

Using simulations, we have compared the finite-sample performance of the estimators with fixed bandwidth. We find that for small bandwidths, the multi-scale estimator is the most precise. The flat-top realized kernel and pre-averaging estimators, however, perform very similarly. In contrast, the non-flat-top realized kernel and modulated realized covariance estimators, which ensure positive semidefiniteness, require much higher bandwidth than the estimators without such a constraint. This is the third contribution of the paper.

Our results can find a use in the implementation of the quadratic covariation estimators in practice. Financial applications include derivative pricing, risk management, portfolio allocation, and high-frequency trading.

\section*{Acknowledgements}
\label{sec:acknow}

We would like to thank the organizers and participants of the 23rd International Conference on Computational Statistics (Iasi, August 28--39, 2018) for fruitful discussions. Computational resources were supplied by the project "e-Infrastruktura CZ" (e-INFRA CZ ID:90140) supported by the Ministry of Education, Youth and Sports of the Czech Republic.

\section*{Funding}
\label{sec:fund}

This research was supported by the Internal Grant Agency of the University of Economics, Prague under project F4/21/2018 and the Czech Science Foundation under project 19-02773S.

%\bibliography{library.bib}
%\bibliographystyle{mynatstyle}

\end{document}